\documentclass[12pt,dvipdfmx]{article}

\renewcommand{\thefootnote}{\fnsymbol{footnote}}

\usepackage{amsbsy,amssymb,latexsym,amsfonts,amsmath}
\usepackage{mathrsfs}
\usepackage[]{graphicx}
\usepackage{bm}
\usepackage{here}
\usepackage{color}

\numberwithin{equation}{section}
\newcommand{\bel}[1]{\begin{equation}\label{#1}}                     
\newcommand{\bal}[1]{\begin{eqnarray}\label{#1}}                     
\newcommand{\be}{\begin{equation}}
\newcommand{\ee}{\end{equation}}

\begin{document}
	%
	%
\begin{titlepage}
		\begin{flushright}
			\normalsize
			~~~~
			NITEP 33\\
			OCU-PHYS 506\\
			September, 2019\\
		\end{flushright}
		
		\vspace{15pt}
		
		\begin{center}
			{\LARGE Multicritical points of unitary matrix model }\\
			\vspace{5pt}
			{\LARGE with logarithmic potential }\\
			\vspace{5pt}
			{\LARGE identified with  Argyres-Douglas points}
		\end{center}
		
		\vspace{23pt}
		
		\begin{center}
			{ H. Itoyama$^{a, b,c}$\footnote{e-mail: itoyama@sci.osaka-cu.ac.jp},
				T. Oota$^{a,b}$\footnote{e-mail: toota@sci.osaka-cu.ac.jp}
				and Katsuya Yano$^b$\footnote{e-mail: yanok@sci.osaka-cu.ac.jp}   }\\
			
			%
			\vspace{10pt}
			%
			
			$^a$\it Nambu Yoichiro Institute of Theoretical and Experimental Physics (NITEP),\\
			Osaka City University\\
			\vspace{5pt}
			
			$^b$\it Department of Mathematics and Physics, Graduate School of Science,\\
			Osaka City University\\
			\vspace{5pt}
			
			$^c$\it Osaka City University Advanced Mathematical Institute (OCAMI)
			
			\vspace{5pt}
			
			3-3-138, Sugimoto, Sumiyoshi-ku, Osaka, 558-8585, Japan \\

		\end{center}
		%
		\vspace{10pt}
		\begin{center}
			Abstract\\
		\end{center}
			In [arXiv:1805.05057 [hep-th]],[arXiv:1812.00811 [hep-th]], the partition function of the Gross-Witten-Wadia unitary matrix model with the logarithmic term has been identified with the $\tau$ function of a certain Painlev\'{e} system, and the double scaling limit of the associated discrete Painlev\'{e} equation to the critical point provides us with the Painlev\'{e} II equation. This limit captures the critical behavior of the $su(2)$, $N_f =2$ $\mathcal{N}=2$ supersymmetric gauge theory around its Argyres-Douglas $4D$ superconformal point. Here, we consider further extension of the model that contains the $k$-th multicritical point and that is to be identified with $\hat{A}_{2k, 2k}$ theory. In the $k=2$ case, we derive a system of two ODEs for the scaling functions to the free energy, the time variable being the scaled total mass and make a consistency check on the spectral curve on this matrix model.

		
		\vfill

\end{titlepage}
	
\renewcommand{\thefootnote}{\arabic{footnote}}
\setcounter{footnote}{0}

\section{Introduction}

In two of the recent publications \cite{IOYanok1,IOYanok2,IOYanokproceedings}, we  pointed out that the double scaling limit \cite{BK,DS,GM} of the zero-dimensional matrix model that was originally introduced long time ago in the context of two dimensional gravity can be successfully exploited to derive the scaling theory in the vicinity of the critical point of Argyres-Douglas type \cite{AD, APSW, EHIY9603,KY} in ${\cal N}=2$ supersymmetric gauge theory. In order to demonstrate this claim, we chose the $su(2)$, $N_f =2$ case that contains the simplest critical point and whose matrix model realization is associated with the $\beta=1$ irregular conformal block \cite{Gai, MMM09092, GT, IOYone, EM09, NR1, NR2, CR, MMS11},
which can be obtained from $2$d conformal block by \cite{DF, MMS10, IO5} the limiting procedure \cite{IOYone}\footnote{See also \cite{MOT1909}.}.
We showed that, in the formulation by the method of orthogonal polynomials \cite{bes79, IZ80, ger54, ger60, sim1, IW0012}, this model itself is exhibited as a certain discrete Painlev\'{e} system \cite{FW, FGR93, NSKGR96}, the partition function being the $\tau$ function of the (continuous) Painlev\'{e} equation \cite{FW02}. (The connection
between the partition function of the matrix models and the $\tau$ function of certain integrable hierarchy
is well known, see for example, \cite{GMMMO91,KMMOZ91}.)

An interesting historical twist is that, while we will not repeat our derivation here, this matrix model turns out to be a unitary (rather than hermitian) matrix model of Gross-Witten-Wadia type \cite{GW80, wad1212, wad80,CIO9305} augmented by the logarithmic term, namely, the case known not to describe
gravitationally dressed unitary $2$d conformal matter field. We have managed to keep the integer coefficient parameter $M$ of the logarithmic term to survive the double scaling limit and derived the Painlev\'{e} II equation (with this parameter\footnote{For the Painlev\'{e} II equation without this parameter, see \cite{PS1, PS2}.}) for the scaling function of the free energy and its $t$(time) derivatives \cite{IOYanok1,IOYanok2,IOYanokproceedings}. Here $t$ and $M$ are the scaling variables associated respectively with the sum and the difference of the two mass parameters.

One of the next several directions of the research that follows from these developments is
to further extend the model to include the series of multicritical points, which we consider in what follows. In the next section, we consider a general single trace symmetric unitary matrix model that contains polynomials up to order $k$ and the inverse and recall its formulation by the orthogonal polynomial. We show that the $k$-th multicritical point is to be identified with the AD point of $\hat{A}_{2k,2k}$ \cite{NR2} gauge theory from the spectral curve.
In section three, we consider the double scaling limit of the model in the $k=2$ case and
derive a system of two ODEs for the scaling functions to the free energy. In section 4, we show that
the double scaled spectral curve for $k=2$ is a mass deformed spectral curve toward
the $(A_1,A_7)$ AD point.

\section{Unitary matrix model and its critical behavior}

Let us consider the unitary matrix model
\begin{align}
	Z := \frac{1}{N !} \left( \prod_{I=1}^N \int d \mu(z_I) \right) \Delta (z) \Delta(1/z), \label{unitarypf}
\end{align}
where $N$ is the size of the unitary matrix, $\Delta(z)$ is the Vandermonde determinant $\Delta(z) = \prod_{I<J} (z_I - z_J)$ and $d \mu(z)$ is
\begin{align}
	\int d\mu(z) := \oint \frac{d z}{2 \pi i z} \exp \left[ -\frac{1}{\underline{g}{}_s} \Bigl( V_k(z) + V_{-k}(z) \Bigr) + M \log z \right], \label{measure}
\end{align}
\begin{align}
	V_k(z) := \sum_{p=1}^k t_p z^p,\qquad V_{-k}(z) := \sum_{p=1}^k t_{-p} z^{-p}. \label{potential}
\end{align}
Here, $\underline{g}{}_s$ is dimensionless parameter related to the dynamical scale $\Lambda$ for $\hat{A}_{2k,2k}$ theory by $\underline{g}{}_s := g_s / \Lambda$\footnote{We have changed the definition of $\underline{g}{}_s$ from \cite{IOYanok1,IOYanok2}, such that the critical value of $\widetilde{S} := \underline{g}{}_s N$ is always $1$. For $k = 1$,  they are related by $2 \underline{g}{}_s^{\tiny \cite{IOYanok1,IOYanok2}} = \underline{g}{}_s / t_1$.}, where $g_s$ is related to the omega background parameters $\epsilon_{1,2}$ as $g_s^2 = - \epsilon_1 \epsilon_2,$ $(\epsilon_2 = -\epsilon_1)$.
We assume that $M$ is an integer.
Note that, up to normalization of the partition function, we can always set the coupling constants $t_k$ and $t_{-k}$ to be identical by the rescaling of the eigenvalues of unitary matrix.
Though the integral contour is also rescaled under this rescaling, it can be deformed to the unit circle since the integrand of eq.\eqref{unitarypf} has no branch cuts.
We thus regard $t_{-k} = t_{k}$ in this paper.

The relevance of the potential of this type to $\hat{A}_{2k,2k}$ theory has been discussed in \cite{NR2}.
Our previous discussion in \cite{IOYanok1,IOYanok2} tells that the appropriate matrix model is not a hermitian matrix model, but a unitary matrix model.
In $k = 1$ case\footnote{$\hat{A}_{2,2}$ theory is the $\mathcal{N} = 2$ $su(2)$ supersymmetric gauge theory with two flavors.}, the matrix model representing the Nekrasov partition function \cite{Nek} is defined by the two integration paths \cite{IOYone}.
In \cite{IOYanok1, IOYanok2}, we have considered the ``generating function" which is defined by a single complex contour and have treated as the unitary matrix model by restricting ourselves to $M \in \mathbb{Z}$.
We assume that the above setup can be generalized to arbitrary $k$.
We thus regard the matrix model related to the $\hat{A}_{2k,2k}$ theory as the unitary matrix model.

\subsection{Orthogonal polynomials, string equations and the $k$-th multicritical point}

To evaluate the partition function \eqref{unitarypf}, let us introduce the orthogonal polynomials
\begin{align}
	p_n(z) := z^n + \ldots + R_n D_n,\qquad \tilde{p}_n(1/z) := z^{-n} + \ldots + \frac{R_n}{D_n}, \label{orthopoly}
\end{align}
where only the zero-th order coefficients are directly related to the partition function.
The orthogonality condition reads
\begin{align}
	\int d \mu(z) \, p_n(z) \tilde{p}_m(1/z) = h_n \delta_{n,m} \, ,
\end{align}
where $h_n$ are the normalization constants. They are related to $R_n$ by
\begin{align}
	\frac{h_n}{h_{n-1}} = 1 - R_n^2. \label{hRrel}
\end{align}
Using this relation \eqref{hRrel}, the partition function \eqref{unitarypf} turns out to be
\begin{align}
	Z = \prod_{k=0}^{N-1} h_k =  h_0^N \prod_{j=1}^{N-1} \left( 1 - R_j{}^2 \right)^{N-j}. \label{pfinRn}
\end{align}

The orthogonal polynomials \eqref{orthopoly} obey
\begin{align}
	z p_n(z) = p_{n+1} (z) + \sum_{k=0}^n C^{(n)}_k \, p_k(z),\qquad 
	z^{-1} \tilde{p}_{n} (1/z) = \tilde{p}_{n+1} (1/z) + \sum_{k=0}^n \widetilde{C}^{(n)}_k \, \tilde{p}_k (1/z), \label{recrelpoly}
\end{align}
where
\begin{align}
C^{(n)}_k =&-  R_{n+1} \left\{ \prod_{j = k+1}^n (1 - R_j{}^2)\right\} R_k \frac{D_{n+1}}{D_k},\\
\widetilde{C}^{(n)}_k =& - R_{n+1} \left\{ \prod_{j = k+1}^n (1 - R_j{}^2)\right\} R_k \frac{D_{k}}{D_{n+1}}.
\end{align}
Using the following identity
\begin{align}
	0 = \int \frac{\partial}{\partial z} \left\{ d \mu(z) z^{k+1} p_\ell(z) \tilde{p}_{m} (1/z) \right\}, \label{identity}
\end{align}
we can obtain the recursion relations for $R_n$ and $D_n$.

In particular, let us consider the cases $(k, \, \ell ,\, m) = (-1 \,, n\, , n-1)$ and $(k, \, \ell ,\, m) = (0 \,, n\, , n)$.
Eq.\eqref{identity} becomes
\begin{align}
	 \int d\mu(z) \Bigl( V_k^\prime(z) + V_{-k}^\prime(1/z) \Bigr) p_n (z) \tilde{p}_{n-1} (1/z) =&\,  -\frac{n}{N} \widetilde{S} \left( h_n - h_{n-1} \right) + \frac{\widetilde{S} \, M}{N} h_n, \label{st1}\\
	\int d\mu(z) z \Bigl( V_k^\prime(z) + V_{-k}^\prime(1/z) \Bigr) p_n (z) \tilde{p_n}(1/z) =& \, \frac{\widetilde{S} \, M}{N} h_n \label{st2},
\end{align}
where $\widetilde{S} := \underline{g}{}_s N$.
Eqs. \eqref{st1} and \eqref{st2} are called the string equation.
After solving these equations for $R_n$, the partition function \eqref{unitarypf} is given by eq.\eqref{pfinRn}.

Note that there is no contribution for $M$ in the large $N$ limit since eqs.\eqref{st1} and \eqref{st2} depend on $M$ at $\mathcal{O}(1/N)$.
Moreover, in the large $N$ limit, these equations reduce to the string equation considered in \cite{PS2}
when $t_{-p} = t_{p}$ for $p = 1,\ldots ,k-1$.
Thus, as in \cite{PS2} the critical value of $t_p$ and $t_{-p}$, denoted by $t^{\ast (k)}_p$, are 
\begin{align}
	t^{\ast (k)}_p =\frac{(-1)^{p+1} \Gamma(k+1)^2}{p \Gamma(k + p +1) \Gamma(k-p+1)}. \label{critical pt}
\end{align}
On this critical point, eq.\eqref{st1} becomes 
\begin{align}
	\widetilde{S} x = 1 - R(x)^{2k},
\end{align}
in the planar limit $N \rightarrow \infty, \,  n/N \rightarrow x$ and $R_n \rightarrow R (x)$.
(Eq.\eqref{st2} becomes trivial.)
The $k$-th multicritical behavior of the planar free energy is
\begin{align}
	F_0 := \lim_{N \rightarrow \infty } - \frac{1}{N^2} \log Z \sim (1 - \widetilde{S})^{\frac{2k + 1}{k}}.
\end{align}
The susceptibility $\gamma$ is equivalent to the no logarithmic case: $\gamma = -1/k$.

\subsection{ Spectral curve }

Let us see that the critical value \eqref{critical pt} corresponds with the Argyres-Douglas point of $\hat{A}_{2k,2k}$ theory by evaluating the spectral curve.
Note that, modulo the problem of integration paths, eq.\eqref{unitarypf} can be rewritten as the hermitian matrix model
\begin{align}
Z =& \frac{1}{N!} \left( \prod_{I=1}^N \int \frac{d z_I}{2 \pi i} \right) \Delta (z)^2 \exp\left[ \frac{1}{g_s} \sum_{I=1}^N W_k(z_I) \right], \label{HUMM}\\
W_k (z) =& -\left( \Lambda V_k(z) + \Lambda V_{-k}(z) - g_s M \log z\right) - S \, \log z\nonumber\\
 =&  -\left( \sum_{\ell=1}^{k} \frac{\mathfrak{t}_\ell}{\ell} z^\ell - \frac{\mathfrak{t}_{-\ell}}{\ell} z^{-\ell} \right) - \mathfrak{t}_0\log z, \label{hpt}
\end{align}
where we define $S := g_s N$, $\mathfrak{t}_0 := S - g_s M$ and  $\mathfrak{t}_\ell := \ell \, t_\ell \Lambda$ for $\ell \ne 0$.
Then we can compute the spectral curve
\begin{align}
y(z)^2 = \left( \frac{W_{\rm pl}^{\prime}{}_k (z)}{2} \right)^2 + f_{\rm pl} (z), \label{curve}
\end{align}
where $W_{\rm pl}{}_k(z)$ and $f_{\rm pl} (z)$ are
\begin{align}
W_{\rm pl}{}_k(z) = \lim_{\substack{N \rightarrow \infty \\ g_s \rightarrow 0}} W_k(z) ,\qquad f_{\rm pl} (z) := \lim_{\substack{N \rightarrow \infty \\ g_s \rightarrow 0}} \left< \!\!\! \left<  g_s \sum_{I=1}^N \frac{W^{\prime}_k (z) - W^{\prime}_k(z_I) }{z - z_I} \right> \!\!\! \right>.
\end{align}
Here, $\left<\!\left< \cdots \right>\!\right>$ means the expectation value with respect to eq.\eqref{HUMM}.
$f_{\rm pl}(z)$ can be written as
\begin{align}
f_{\rm pl}(z) = \frac{1}{z^2}\sum_{\ell = -k+1}^k d_\ell \, z^\ell,
\end{align}
where $d_\ell$ is a function of $\mathfrak{t}^{(k)}_\ell$ and 
\begin{align}
	p_\ell = \lim_{\substack{N \rightarrow \infty \\ g_s \rightarrow 0}} \left< \! \!\!\left<  g_s \sum_{I=1}^N w_I^\ell \right>\! \!\! \right>.
\end{align}
Then eq.\eqref{curve} becomes
\begin{align}
y(z)^2 =& \frac{1}{4 z^2} \left\{ \mathfrak{t}_k^2 \, z^{2k} +  \sum_{\ell = k}^{2k-1} \left( - 4 \mathfrak{t}_k S\, \delta_{\ell,k} + \sum_{p = \ell - k}^k \mathfrak{t}_p \mathfrak{t}_{\ell - p}  \right) z^\ell + \sum_{\ell = - k + 1}^{k-1} \left( - 4 d_\ell + \sum_{p + q = \ell} \mathfrak{t}_p \mathfrak{t}_q \right) z^\ell \right.\nonumber\\
&\left. + \sum_{\ell = k}^{2k} \left( \sum_{p = \ell - k}^k \mathfrak{t}_{-p} \mathfrak{t}_{-(\ell - p)} \right) z^{-\ell} + \frac{\mathfrak{t}_k^2}{z^{2k}} \right\}. \label{sc}
\end{align}
In eq.\eqref{sc}, we have used $\mathfrak{t}_{-k} = - \mathfrak{t}_k$.
After the renormalizing the dynamical scale by $\Lambda \rightarrow 2 \Lambda/ ( k \,t_k)$,
the spectral curve \eqref{sc} takes the same form as the Seiberg-Witten curve of $\hat{A}_{2k,2k}$ theory \cite{NR2}.

On the critical point \eqref{critical pt}, the spectral curve \eqref{sc} degenerates to
\begin{align}
	y(z)^2 = \frac{\Lambda^2}{z^{2k +2}} \left( z - 1 \right)^{4k},
\end{align}
if we set $d_\ell$ appropriately.
Then we conclude that the Argyres-Douglas point of $\hat{A}_{2k,2k}$ theory corresponds to the $k$-th multicritical point of the unitary matrix model with the potential \eqref{potential}.

\section{The double scaling limit for $k = 2$ case}
In this section, we consider the potential
\begin{align}
	-\frac{1}{\underline{g}{}_s} \Bigl( V_2 (z) + V_{-2}(z) \Bigr) + M \log z = -\frac{N}{\widetilde{S}} \left\{ t_2 \left( z^2 + \frac{1}{z^2} \right) + t_1 z + t_{-1}\frac{1}{z} \right\} + M \log z.
\end{align}
In order to study the $k = 2$ multicritical behavior, we should take the double scaling limit of the string equations.
To do this, let us write eqs.\eqref{st1} and \eqref{st2} for $k=2$
\begin{align}
	\frac{n}{N} \widetilde{S} \frac{R_n^2}{1 - R_n^2} =& \,\,  t_{-1} \left( R_{n+1} R_n \frac{D_n}{D_{n+1}} + R_{n} R_{n-1} \frac{D_{n-1}}{D_n} \right) - 2 t_2 \left\{ R_{n+1}R_{n-1} \frac{D_{n+1}}{D_{n-1}} \right. \nonumber\\
	&\left. - R_{n+2} \left( 1 - R_{n+1}^2 \right) R_n \frac{D_n}{D_{n+2}} - R_{n+1} \left( 1 - R_{n}^2 \right) R_{n-1} \frac{D_{n-1}}{D_{n+1}}  \right.\nonumber\\
	&\quad - R_{n} \left( 1 - R_{n-1}^2 \right) R_{n-2} \frac{D_{n-2}}{D_{n}} + \left. R_{n+1}^2 R_{n}^2 \frac{D_n^2}{D_{n+1}^2} + R_{n+1} R_n^2 R_{n-1} \frac{D_{n-1}}{D_{n+1}} \right. \nonumber\\
	&\qquad\left. + R_{n}^2 R_{n-1}^2 \frac{D_{n-1}^2}{D_{n}^2} \right\} - \frac{M \widetilde{S}}{N}, \label{k=2 st1}\\
	0 =& \,\, R_{n+1} R_n \left( t_1 \frac{D_{n+1}}{D_n} - t_{-1} \frac{D_n}{D_{n+1}} \right) +2 t_2 \left\{ R_{n+2} \left( 1 - R_{n+1}^2 \right) R_n \left(  \frac{D_{n+2}}{D_n} -  \frac{D_n}{D_{n+2}} \right) \right. \nonumber\\
	&\left. + R_{n+1} \left( 1 - R_{n}^2 \right) R_{n-1} \left(  \frac{D_{n+1}}{D_{n-1}} -  \frac{D_{n-1}}{D_{n+1}} \right) -  R_{n+1}^2 R_n^2 \left(  \frac{D_{n+1}^2}{D_n^2} -  \frac{D_n^2}{D_{n+1}^2} \right) \right\} \nonumber\\
	&\quad + \frac{M \widetilde{S}}{N}  \label{k=2 st2}
\end{align}
When $M = 0$ and $t_1 = t_{-1}$, eq.\eqref{k=2 st2} becomes trivial and eq.\eqref{k=2 st1} reduce to the string equation considered in \cite{PS1,PS2}.

To take the double scaling limit, let us define the scaling ansatz as $x := n /N, \, a^5 = 1/N, \, \widetilde{S} x = 1 - a^4 t$ and
\begin{align}
	R_n \equiv R \left( \frac{n}{N}  \right) = R(x) = a u (t), \qquad D_n \equiv D \left( \frac{n}{N} \right) = D(x) = d(t). \label{dsl}
\end{align}
With these ansatz, the double scaling limit is defined as the $a \rightarrow 0$ while keeping
\begin{align}
	\kappa \equiv  \frac{a^5}{(1 - \widetilde{S})^{5/4}}, \label{dresscpl}
\end{align}
finite.
Note that, in this limit, $R_{n + k}$ becomes
\begin{align}
	R_{n + k} = R \left( \frac{n}{N} + \frac{k}{N} \right) = R (x +  a^5 k) =& \sum_{n = 0}^\infty \frac{(a^5 k)^n}{n !} \frac{d^n}{ d x^n} R(x)\nonumber \\ 
	=& \sum_{n = 0}^\infty \frac{a^{n + 1} k^n}{n!} \left\{ - (1 + \tilde{c} a^4) \right\}^n \frac{d^n}{d t^n} u(t),
\end{align}
where we used
\begin{align}
\frac{d}{dx} = \frac{d t}{d x} \frac{d}{dt} = - a^{-4} (1 +  \tilde{c} a^4) \frac{d}{dt}, \qquad \tilde{c} := -\kappa^{-4/5}.
\end{align}
Similarly, $D_{n + k}$ takes
\begin{align}
	D_{n + k} = \sum_{n = 0}^\infty \frac{a^{n} k^n}{n!} \left\{ - (1 + \tilde{c} a^4) \right\}^n \frac{d^n}{d t^n} d(t).
\end{align}

Setting the couplings at them critical value $t_{\pm1} = t^{\ast (2)}_1 = 2/3 , \, t_2 = t^{\ast (2)}_2 = -1/12$ and substituting scaling ansatz defined above into the eqs.\eqref{k=2 st1} and \eqref{k=2 st2}, we obtain
\begin{align}
	0 =& \frac{u^2}{3}\left\{ \frac{d^{(3)}}{d} - 3 \frac{d^{(1)} d^{(2)}}{d^2} + 4 \left(\frac{d^{(1)}}{d}\right)^3 \right\} + \frac{2}{3} u u^{(1)} \left\{ \frac{d^{(2)}}{d} - \left(\frac{d^{(1)}}{d} \right)^2 \right\} \nonumber\\
	 &+ \left( \frac{4}{3} u u^{(2)} - \frac{2}{3} u^{(1)}{}^2 - 2 u^4 \right) \frac{d^{(1)}}{d} + M, \label{dslstring1}
\end{align}
from eq.\eqref{k=2 st2} at $\mathcal{O}(a^5)$.
From eq.\eqref{k=2 st1}, we also get the same equation as eq.\eqref{dslstring1} at $\mathcal{O}(a^3)$ and have
\begin{align}
	u t =& \frac{u^{(4)}}{6} - \frac{5}{3} u^2 u^{(2)} - \frac{5}{3} u u^{(1)}{}^2 + u^5 + \left(  u^{(2)} - 2u^3\right) \left( \frac{d^{(1)}}{d} \right)^2 + 2 u^{(1)} \left\{ \frac{d^{(1)} d^{(2)}}{d^2} -\left( \frac{d^{(1)}}{d} \right)^3 \right\} \nonumber\\
	&+ 6u \left\{ 4 \frac{d^{(1)} d^{(3)}}{d^2} + 3 \left( \frac{d^{(2)}}{d} \right)^2 - 18 \frac{d^{(1)}{}^2 d^{(2)} }{d^3} + 12 \frac{d^{(1)}{}^4}{d^4}\right\}, \label{dslstring2}
\end{align}
at $\mathcal{O}(a^4)$.
Here, $u = u(t)$, $d = d(t)$ and
\begin{align}
	u^{(n)} = \frac{d^n}{d t^n} u(t),\qquad d^{(n)} = \frac{d^n}{d t^n} d(t).
\end{align}
After eliminating $d^{(3)}$ from eq.\eqref{dslstring2} by using eq.\eqref{dslstring1}, we obtain
\begin{align}
	u t =& \frac{u^{(4)}}{6} - \frac{5}{3} u^2 u^{(2)} - \frac{5}{3} u u^{(1)}{}^2 + u^5 - \frac{2 M}{u} \frac{d^{(1)}}{d} \nonumber\\
	&+ \left\{ 2 u^3 + \frac{4 u^{(1)}{}^2}{3 u} - \frac{5}{3} u^{(2)} \right\} \left(\frac{d^{(1)}}{d}\right)^2 + \frac{2 u^{(1)}}{3} \left\{ \frac{d^{(1)} d^{(2)}}{d^2}  - \left( \frac{d^{(1)}}{d} \right)^3 \right\} \nonumber\\
	&\quad -\frac{u}{6} \left\{ 3 \left( \frac{d^{(2)}}{d} \right)^2 - 6 \frac{d^{(1)}{}^2 d^{(2)} }{d^3} - 4 \left( \frac{d^{(1)}}{d} \right)^4 \right\}. \label{dslstring3}
\end{align}
Defining $g = d^{(1)} / d$, eqs.\eqref{dslstring1} and \eqref{dslstring3} can be written by
\begin{align}
	u\,t =&  \frac{u^{(4)}}{6} - \frac{5}{3} u^2 u^{(2)} - \frac{5}{3} u u^{(1)}{}^2 + u^5  - 2 M \frac{g}{u} \nonumber\\
	&+ \left( 2 u^3 + \frac{4 u^{(1)}{}^2}{3 u} - \frac{5}{3} u^{(2)} \right) g^2 + \frac{2}{3} u^{(1)} g g^{(1)}  - \frac{u}{6} \left( 3 g^{(1)} - 7 g \right), \label{dslstr1}  \\
	0 =& \frac{u^2}{3} \left( g^{(2)} + 2 g^3 \right) + \frac{2}{3} u u^{(1)} g^{(1)} + \left( \frac{4}{3} u u^{(2)} - \frac{2}{3} u^{(1)}{}^2 - 2 u^4 \right) g + M. \label{dslstr2}
\end{align}
Hence, we obtain two differential equations which characterize the partition function of our matrix model in the double scaling limit.
Appearance of the two string equations is known in the hermitian matrix model with general potential \cite{HMPN}. We expect that eqs.\eqref{dslstr1} and \eqref{dslstr2} are the unitary matrix model version.
If we consider the $M=0, \, t_1 = t_{-1}$ case, $D_n$ becomes $1$ for all $n$. Then, eqs.\eqref{dslstr1} and \eqref{dslstr2} reduce to the fourth order differential equation for $u(t)$ obtained in \cite{PS1, PS2}.

\section{The corresponding Argyres-Douglas theory}
In this section, we take the same limit in the last section to the spectral curve.
For $k = 2$ case, the spectral curve \eqref{sc} takes the following form:
\begin{align}
	y(z)^2 = \frac{1 }{4 z^6} \left( \sum_{i = 0}^8 c_i z^i \right), \label{22curve}
\end{align}
where $c_0 =  c_8 = \mathfrak{t}_2{}^2$,
\begin{equation}
\begin{split}
	c_1 =& 2\,\mathfrak{t}_{-2} \mathfrak{t}_{-2},\qquad c_2 = \mathfrak{t}_{-1}{}^2 + 2\, \mathfrak{t}_{-2} \mathfrak{t}_0,\qquad c_3 = 4 p_{-1} \mathfrak{t}_{-2} + 2 \mathfrak{t}_{-1} \mathfrak{t}_{0} + 2 \mathfrak{t}_{-2} \mathfrak{t}_1,\\
	c_4 =& 4 p_{-2} \mathfrak{t}_{-2} + 4p_{-1} \mathfrak{t}_{-1} + \mathfrak{t}_0{}^2 + 2 \mathfrak{t}_{-1} \mathfrak{t}_1 + 2 \mathfrak{t}_{-2} \mathfrak{t}_{2},\\
	c_5 =& -4 p_1 \mathfrak{t}_{2}  +2( \mathfrak{t}_0 - 2 S ) \mathfrak{t}_1 + 2 \mathfrak{t}_{-1} \mathfrak{t}_2, \qquad
	c_6 = \mathfrak{t}_1{}^2 + 2 ( \mathfrak{t}_0 - 2 S ) \mathfrak{t}_0 ,\qquad c_7 = 2 \mathfrak{t}_1 \mathfrak{t}_2.
\end{split}
\end{equation}
At the $k = 2$ multicritical point
\begin{align}
	S^\ast = \Lambda,\qquad \mathfrak{t}_0^\ast = S^\ast  =  \Lambda,\qquad \mathfrak{t}_{\pm1}^\ast = \pm \frac{2}{3} \Lambda, \qquad \mathfrak{t}_{\pm 2}^\ast = \mp \frac{1}{6} \Lambda,
\end{align}
the curve \eqref{22curve} becomes
\begin{align}
	y(z)^2 = \frac{\Lambda^2}{144 z^6} \left( 1 - z \right)^8,
\end{align}
at
\begin{align}
	p_{1} = p_{-1} = - \frac{2}{3} \Lambda,\qquad p_{-2} = \frac{1}{6} \Lambda.
\end{align}

Let us take the scaling limit defined by eqs.\eqref{dsl} and \eqref{dresscpl}, namely
\begin{align}
	S =& \Lambda + a^4 \tilde{c}\Lambda  ,\\
	\mathfrak{t}_0 =& S - \frac{S \, M}{N} =  \Lambda + a^4 \tilde{c} \Lambda - a^5 M \Lambda + \mathcal{O}(a^9).
\end{align}
We also define the following scaling ansatz for $z$, $p_{\pm1}$ and $p_{-2}$:
\begin{align}
	z =& 1 +  a \tilde{z},\\
	p_1 =& -\frac{2}{3}\Lambda + a^5 M \Lambda  +  a^6 \tilde{u}_1 \Lambda  + a^7 \tilde{u}_2 \Lambda ,\\
	p_{-1} =&-\frac{2}{3}\Lambda - a^5 M \Lambda  + a^6 \tilde{u}_1 \Lambda  - a^7  \tilde{u}_2 \Lambda ,\\
	p_{-2} =& \frac{1}{6} \Lambda - a^5 M \Lambda  + 2 a^6  \tilde{u}_1 \Lambda  - 4 a^7 \tilde{u}_2 \Lambda  + a^8 \tilde{u}_3 \Lambda .
\end{align}
Substituting these scaling ansatz into \eqref{22curve}, we obtain
\begin{align}
	y(z)^2 = \frac{ a^8 \Lambda^2  }{144} \left( \tilde{z}^8 +12 \tilde{c} \tilde{z}^4 + 24 M \tilde{z}^3 + 24 \tilde{u}_1 \tilde{z}^2 +48 \tilde{u}_2 \tilde{z} + 36 \tilde{c}^2 + 24 \tilde{u}_3 \right) + \mathcal{O}(a^9).
\end{align}
We thus define $\tilde{y} = 12 y / ( a^4 \Lambda ) $, then the curve \eqref{22curve} turns into
\begin{align}
	\tilde{y}^2 = \tilde{z}^8 +12 \tilde{c} \tilde{z}^4 + 24 M \tilde{z}^3 - 24 \tilde{u}_1 \tilde{z}^2 +48 \tilde{u}_2 \tilde{z} + 36 \tilde{c}^2 + 24 \tilde{u}_3. \label{A1A7AD}
\end{align}
in $a \rightarrow 0$ limit.
This curve \eqref{A1A7AD} is the same form of the $(A_1, A_7)$ Argyres-Douglas theory \cite{DX} except the sixth order and fifth order term in $\tilde{z}$.

The scaling dimension for $\tilde{y}$ and $\tilde{z}$ are
\begin{align}
	[ \tilde{y} ] = \frac{4}{5},\qquad [ \tilde{z} ] = \frac{1}{5}.
\end{align}
Hence, scaling dimensions of various parameters in the curve \eqref{A1A7AD} are
\begin{align}
	[\tilde{c}] = \frac{4}{5},\qquad [M] = 1,\qquad [\tilde{u}_i] = \frac{5 + i}{5}.
\end{align}
Therefore, we can see that the parameter $\tilde{c}$ fixed in the double scaling limit corresponds the relevant parameter in the $(A_1 , A_7)$ Argyres-Douglas theory.
The coupling constant of logarithmic term $M$ plays the role of the mass parameter in $(A_1 , A_7)$ Argyres-Douglas theory.
If we scale also the coupling constants $t_{\pm 1}$ from their critical value, we expect that the full parameters in the curve are recovered.

\section*{Acknowledgment}

The work of H.I. is partly supported by JSPS KAKENHI Grant Number 19K03828 and OCAMI MEXT Joint Usage/Research Center on Mathematics and Theoretical Physics.

\end{document}